\documentclass[pre,aps,
notitlepage,
nofootinbib,
superscriptaddress,floats,floatfix,showpacs,
pdflatex10pt]{revtex4}
\usepackage{graphicx}
\usepackage{dcolumn}
\usepackage{bm}
\usepackage{amssymb}
\usepackage{amsmath}
\usepackage{mathrsfs}
\usepackage{ulem}
\usepackage{color}

\newcommand{\re}[1]{{\color{red}#1}}

\begin{document}
\title{Nonlinear waves in a dispersive vacuum described with   a
high order derivative electromagnetic Lagrangian}
\author{Francesco Pegoraro}
\affiliation{Enrico Fermi Department of Physics, University of Pisa, 56127 Pisa, Italy}
\affiliation{ National Research Council, National Institute of Optics, via G. Moruzzi 1, 56124 Pisa, Italy}
\author{Sergei V. Bulanov}
\affiliation{Institute of Physics of the ASCR, ELI--Beamlines project, Na Slovance 2, 18221 Prague, Czech Republic}
\affiliation{National institutes for Quantum and Radiological Science and Technology (QST),
Kansai Photon Science Institute, 8--1--7 Umemidai, Kizugawa, 619--0215 Kyoto, Japan}


\date{\today}
\begin{abstract}
In this article we use an  electromagnetic Lagrangian constructed  so as to include 
dispersive effects in  the description of an electromagnetic wave  propagating 
in the Quantum Electrodynamic Vacuum. 
This  Lagrangian is Lorentz invariant, includes contributions up to six powers 
in  the electromagnetic fields and  involves both   fields and  their  first derivatives.   
Conceptual  limitations inherent to the use of this higher derivative Lagrangian approach  
are   discussed.   
We consider  the one-dimensional spatial limit and obtain  an exact solution of the nonlinear  
wave equation recovering the Korteveg-de Vries type periodic waves and solitons  
 given in  S. V. Bulanov {\it et al.},  {\it Phys. Rev. D}, {\bf  101}, 016016 (2020).
 \end{abstract}
\maketitle

 \section{Introduction:  wave equations  in nonlinear quantum electrodynamics}
 \label{intro}

Field induced polarization and birefringence  of the vacuum, see e.g. Refs.\cite{BLP-QED, VLG}, 
are  fundamental effects predicted by Quantum electrodynamics (QED).  
These effects arise from the process of scattering of light by light: while 
in classical electrodynamics electromagnetic waves do not interact in vacuum,  
in QED photon-photon scattering can take place in vacuum via the generation 
of virtual electron-positron pairs that gives rise to polarization 
and magnetization currents that make the vacuum respond as a material medium.  
The study of the nonlinear QED vacuum properties has been conducted for about a century 
\cite{HeisenbergEuler, Karplus1951, Schw51, McK1963, Erber1966, Mamaev, Grib1980, Bernard2000, Batt2018}.

Recently, interest in these effects have been rekindled  by 
the availability of high power lasers (see review articles \cite{CDanson2015, CDanson2019}) 
leading to the formulation of the plans aimed at reaching experimentally the parameters that can enable the study
the nonlinear QED vacuum \cite{Heinzl2006, Schlenvoigt2016, King}. 
This, in its turn,  has motivated an intensive   theoretical  research  program aimed at the study of  the scattering 
of a laser pulse by a laser 
pulse \cite{Mourou, Marklund, DTom, Pare, DiPiazzaReview, BattRizz, Koga, KarbsteinShai1, KarbsteinShai2, Angioi2019, SVBRFAL2019, ZHANG}. 

The field induced  vacuum polarization and birefringence can be accounted for  
within the framework of a local approximation 
using the well known Heisenberg-Euler Lagrangian in the electromagnetic action functional 
\cite{HeisenbergEuler, Schw51}.  
This  approximation   leads to nonlinear wave equations for  the fields amplitude in vacuum  that are not dispersive, 
i.e. that are homogeneous 
in the second order derivatives of the field four-vector potentials. In  other words the Lagrangian does
 not include  second order  derivatives 
(or higher order derivatives) of the electromagnetic fields. This local approximation is valid in the long wavelength, 
low frequency limit, essentially requiring that the electromagnetic fields are slowly varying 
on the Compton scattering wavelength $ \lambdabar_C=\hbar/m_ec$,  
where $\hbar$ is the reduced Planck constant, $e$ and $m_e$ are the electron electric charge 
and mass and $c$ is the speed of light in vacuum.  

Non local effects on the Compton scattering have been studied e.g. in Ref. \cite{ADP2018} and, 
for vacuum birefringence, in Ref. \cite{FK2015}.

For shorter wavelengths the vacuum acquires dispersion properties. In the small field amplitude limit and 
in the so called cross  field approximation these dispersive properties have been included 
in the ``invariant photon mass'' introduced in Ref.  \cite{Ritus} 
 (see also   Refs. \cite{Narozhnyi1969, Ritus1972, Karb3, Adler1971, PRDS}). 
 The cross field approximation consists in approximating the interaction between 
 a higher frequency pulse and a lower frequency pulse 
by taking the latter to be described by uniform and stationary electric and magnetic fields 
of the same amplitude and orthogonal to each other.  Discussions of the QED processes 
beyond the constant field approximations 
can be found in Ref. \cite{Tamb, DiPiazza2019, Ilderton2019a, Ilderton2019b}.
The invariant photon mass refers to the higher frequency pulse, it depends on the relative polarization 
between the fields of the two pulses and  can be expressed  \cite{Ritus} in terms 
of the so called quantum nonlinearity parameter $\chi_{\gamma}$.
This Lorentz invariant parameter  can be written for a photon impinging on an  slowly varying external field as
\cite{DiPiazzaReview, Ritus}
\begin{equation}
\label{chi}
 \chi_{\gamma} = \frac{e}{m_e^3}\, \sqrt{ -(F_{\mu\nu} k^\nu)^2 }\, ,
\end{equation}
where $k^\nu$ is the 4-wave {vector} on the impinging photon and $ F_{\mu\nu}$  
the electromagnetic external  field tensor. Natural units  are adopted with   $c=\hbar=1$,
 the fields in the electromagnetic field tensor are normalized on the critical QED field $E_S $ 
 (which is given in dimensional units by $E_S =m_e^2c^3/e\hbar$ \cite{SAU31}). 
In the limit of small $ \chi_{\gamma}$, for parallel polarizations the  
square of the photon invariant mass  ${\mathfrak{m}}$ can be  written as 
\begin{equation}
\label{mass}
{\mathfrak{m}}^2 = -\alpha m_e^2 \left\{\frac{4}{45 \pi}\left[\chi_{\gamma}^2+\frac{1}{3}\chi_{\gamma}^4\right]
+i\frac{1}{4}\sqrt{\frac{3}{4}}\chi_{\gamma}\exp{\left(-\frac{2}{3\chi_{\gamma}}\right)}
+{\cal O}\left(\chi_{\gamma}^6\right)\right\} \, ,
\end{equation}
where $\alpha = e^2/\hbar c$ is the fine structure constant. The imaginary part is exponentially small. The term 
proportional to {$\chi_{\gamma}^4$} describes the dispersion effects, 
i. e. the effects corresponding to the wave propagation 
velocity dependence on the wave vector.
This result was used  (to leading order in $\chi_{\gamma}^4$)  in Ref. \cite{PRDS}  to derive 
a nonlinear wave equation   
for finite amplitude dispersive waves counterpropagating with respect to a cross field configuration.  This wave 
equation is of the form of the Korteveg-De Vries (KdV) equation \cite{KdV1885} 
in the one-dimensional spatial case and 
 includes {third} order derivatives of the impinging wave vector potential while it has the form of
   the Kadomtsev-Petviashivili equation 
\cite{KP1970, KP2018} in the two-dimensional case. In one-dimensional case, in light-cone coordinates 
$x^+, x^-$ 
and for normalized   variables (for explicit  definitions see Sec. \ref{basic}), 
the KdV equation obtained in Ref. \cite{PRDS}   describing the 
electromagnetic wave in the QED vacuum reads
\begin{equation}
\label{1}
\partial_{+} a-\left( \kappa_{1}+\kappa_{2}a\right)\partial_{-} a - 2 \kappa_{3}\partial_{---} a = 0,
\end{equation}
where the coefficients  $\kappa_1 = 4\alpha W_0^2/(45 \pi)$,  $\kappa_2 =32 {\sqrt 2} \alpha W_0^3/(105 \pi)$ and $\kappa_3 =4\alpha W_0^4/(135 \pi)$  are proportional  respectively to the second, 
to the third and to the fourth power of the cross field amplitude $W_o$. The  nonlinear term in   Eq.(\ref{1})  arises
 from the  Heisenberg-Euler Lagrangian \cite{HeisenbergEuler}, truncated at the six photon contribution,
 and  allows  for  propagating Korteveg-de Vries soliton solutions.

In the present article  we  use  a  Lagrangian in the electromagnetic vacuum action  
that involves  higher order derivatives of 
the wave vector potential and that is constructed so as to include the quantum 
nonlinearity parameter dependency of the invariant photon mass. 
 In this formulation higher order derivatives enter in combination  with nonlinear terms. We derive the corresponding 
 field equations by a variation of the action integral and obtain scattering solutions for counter-streaming finite length 
 pulses that include the effect of dispersion.  In addition we derive the general solution for  finite amplitude 
 waves propagating in a  cross field configuration.
These solutions describe a class  of soliton solutions of the type described in Ref. \cite{PRDS}.
\medskip

Effective electromagnetic Lagrangians depending on higher derivatives have been introduced 
in the context of  modified linear electrodynamics,  
 or limited to leading order in the field amplitude and field derivatives  separately,   
 by B. Podolsky \cite{Pod1,Pod2,Pod3},  
 Barut \& Mullen \cite{BM},  Lee \& Wick  \cite{LW}, see  also  Refs. \cite{TN, BBMB, MSMS}. 
 Essentially these Lagrangians are of the form  exemplified by the Lee-Wick Lagrangian 
\begin{equation} \label{LW1}
\mathcal{L} = \mathcal{L}_{0} + \mathcal{L}_{LW},\quad {\rm with } \quad \mathcal{L}_{0}=
-\frac{1}{4\pi}F_{\mu \nu}F^{\mu \nu} \quad {\rm and } \quad  \mathcal{L}_{LW} = \frac{1}{4 M^2} F_{\mu \nu}
(\partial^\alpha\partial_\alpha F^{\mu\nu}),
\end{equation}
 with the  inclusion, see e.g. Ref.\cite{MSMS}, of  the first nonlinear contributions  
 from the Heisenberg-Euler Lagrangian. In Eq.(\ref{LW1})  
 $\mathcal{L}_{0}$ is the classical electromagnetic Lagrangian in vacuum and $M$ is a mass parameter.
We note that a Lagrangian depending on  higher order  derivatives of the form 
 \begin{equation}
 \label{L:MME}
 \mathcal{L}_{MME}=\frac{\alpha}{m_e^2}
 \left[  -\left(\partial_{\kappa} F_{\lambda}^{\kappa} \right)\left(\partial_{\mu} 
 F^{\mu \nu} \right)+F_{\mu \nu}\partial_{\lambda}\partial^{\lambda} F^{\mu \nu}  \right]
  \end{equation}
was obtained in Refs. \cite{Mamaev, Grib1980}.
A derivative expansion of the effective action for nonlinear  quantum electrodynamics  
has been obtained in Refs. \cite{Gusynin1,Gusynin2}
in terms of a Lagrangian that is  written in the form
\begin{equation} 
\label{Gus}
\mathcal{L} = \mathcal{L}_{HE} + \partial_\lambda  F_{\alpha\beta}\partial_\gamma  
F_{\sigma\delta}L_{1}^{\lambda\alpha\beta\gamma  \sigma\delta} (F^{\mu\nu}) 
+  {\rm higher ~field ~ derivative ~ terms},
\end{equation}
where $\mathcal{L}_{HE}$  is the  Heisenberg-Euler Lagrangian  and  
$L_{1}^{\lambda\alpha\beta\gamma  \sigma\delta}(F^{\mu\nu}) $ is a local function of the electromagnetic field tensor.

\subsection{Well posedness of higher order Lagrangians}

The physical interpretation of higher order derivative Lagrangians  presents 
some difficulties as these Lagrangians   lead to ``ghost'' degrees of freedom  and to instabilities.
In 1850   Ostrogradski \cite{ostr} proved in the context of Classical Mechanics  that a Lagrangian 
of the form ${\cal  L}(q,\dot q,  \ddot q) $,  which requires 4 initial conditions 
and  thus involves 4 canonical variables,  leads to a Hamiltonian  that is not bounded from below with 
respect to a ``ghost'' degree of freedom,   see Ref.\cite{t-jC} where 
it was shown  that in a non-degenerate higher derivative theory, 
that is in a theory where the higher order derivative does not simply  amounts to a total time derivative,  
the Ostrogradski instability can only be removed by the addition of constraints that reduce the  phase space 
of the original theory.

For  systems  with infinite degrees of freedom, higher order derivative  Lagrangians can lead to additional 
wave branches corresponding, 
e.g. for the Lee-Wick  Lagrangian, to two independent (on-shell) spin-1 fields:  the original  massless photon field  
and and an additional  massive one.

In the case of the Podolsky electrodynamics \cite{Pod1, Pod2,Pod3}, as well as in  the case considered 
by Mamaev, Mostepanenko and Eidis \cite{Mamaev}, 
which was introduced in order to regularize the electromagnetic field behaviour at short distances
and to take into account the field inhomogeneity effects,
the Lagrangian can be written as 
 \begin{equation}
 \label{L:POD}
 \mathcal{L}= \mathcal{L}_0+ \mathcal{L}_{Pod}, \quad {\rm with} \quad 
 \mathcal{L}_{Pod}=\frac{1}{M^2}
 \left[  -\left(\partial_{\kappa} F_{\lambda}^{\kappa} \right)\left(\partial_{\mu} 
 F^{\mu \nu} \right)+F_{\mu \nu}\partial_{\lambda}\partial^{\lambda} F^{\mu \nu}  \right] ,
  \end{equation}
where  the inverse mass $1/M$ is the parameter that corresponds to the length, and the current source term
 is not included. From this Lagrangian we obtain in the Lorentz gauge $\partial_{\mu}A^{\mu}=0$ the wave equation
 \begin{equation}
 \label{Eq:WEQ}
\left(1-\frac{1}{M^2}\partial_{\mu} \partial^{\mu}\right)\partial_{\nu}\partial^{\nu}A^{\kappa}=0.
  \end{equation}
It corresponds to a photon branch with dispersion equation $k_{\nu}k^{\nu}=0$ and 
a ghost branch with dispersion equation $k_{\mu}k^{\mu}=-M^2$. The addition of nonlinear terms 
in the Lagrangian will in general couple the two different branches (see Appendix \ref{GHOSTS}). An analogous result can be derived 
from the Lagrangians introduced in \cite{BM, LW}.

For the sake of consistency, see also Ref. \cite{effective}, in  what follows  we will treat the terms with the higher  derivatives as  corrections 
to the classical electromagnetic Lagrangian density $\mathcal{L}_{0}$  and in particular we will  require that any effective mass arising from the balance between the nonlinear and the dispersive terms remains finite in the limit in which the value of the fine structure constant $\alpha$ is set equal to zero.
 
 \subsection{Outline of the article}
 \label{out} 
 
 In Sec.\ref{genlagr}, after specifying  for the sake of clarity some normalization conditions,  
 the structure of the  dispersive Lagrangian term is formalized for an electromagnetic vacuum configuration  
 in 4-dimensional  Minkowski space. Its   reduced expression in a spatially one dimensional configuration 
 in then expressed in terms of light-cone variables.
In Sec.\ref{basic}  the  electromagnetic  field equations are derived from the full one dimensional  
Lagrangian in vacuum including the Heisenberg-Euler
and the dispersion terms up to the sixth power of the  fields.
Then  explicit solutions are obtained  for the scattering 
of two counter-propagating finite length pulses 
Sec.\ref{scatter},  and  for  finite amplitude waves in a constant cross 
field configuration Sec.\ref{SolCon}. These sections are supplemented by  four Appendices at the 
end of the article. The first two Appendices  serve 
the purpose of providing algebraic developments  separately so as not to interrupt the flow of  the presentation.
In the  final  two Appendices  ghost solutions are described in the case of a  nonlinear wave equation with 
linear dispersion terms derived from the Lagrangian given by Eq.(\ref{L:MME}),  Appendix \ref{GHOSTS},   
and of  the  wave equation
derived from the Lagrangian given by Eq.(\ref{Full-Lagrangian0}) for a spacial class of selfsimilar solutions, 
Appendix \ref{A3}.  
 Finally in Sec.\ref{concl} conclusions are drawn and a  possible inclusion  of  higher powers
 of the field amplitudes  in the  Lagrangian  is indicated as a possible path towards the identification 
 of higher order solitonic structures in the process of light-light interaction.
 
\section{Dispersive contribution to the Lagrangian density}
\label{genlagr}

We specify the normalizations adopted in  this article by writing the classical electrodynamics Lagrangian in the form
\begin{equation}\label{Lag1}
\mathcal{L}_{0}=-\frac{m_e^4}{16\pi\alpha}F_{\mu \nu}F^{\mu \nu} = -\frac{m_e^4}{4\pi\alpha}  \, {\mathfrak F}  ,
\end{equation}
while the Heisenberg-Euler  Lagrangian density  (truncated  at the 6-photon interaction term)  is written as 
\begin{equation}
\label{Lag2}
{\mathcal{L}}_{HE}=-\frac{m_e^4 \, e^4}{90\pi^2}  \left[ ({\mathfrak F}^2 + \frac{7}{4} {\mathfrak G}^2) 
+ \frac{8}{7} {\mathfrak F} ({\mathfrak F}^2 + \frac{13}{16} {\mathfrak G}^2)\right] .
\end{equation}
The Lorentz invariants ${\mathfrak F}$ and ${\mathfrak G}$ are defined by 
\begin{equation}
\label{Lag3}
{\mathfrak F}=\frac{1}{4}F_{\mu \nu}F^{\mu \nu},
\quad 
{\mathfrak G}=-\frac{1}{8}\varepsilon^{\mu \nu \kappa \lambda}F_{\kappa \lambda} F_{\mu \nu}, 
\end{equation}
 with $ \varepsilon^{\mu \nu \kappa \lambda}F_{\kappa \lambda} $   the   dual   electromagnetic  field  tensor. 
Here $ \varepsilon^{\mu \nu \kappa \lambda}$ is the fully antisymetric 4 dimensional Levi-Civita tensor.

Referring  to the  quantum nonlinearity parameter $\chi_\gamma$  that is used in the definition 
of the invariant photon mass and in view Eq.(\ref{Gus}), we define the following dispersive 
contribution to the vacuum Lagrangian density
 \begin{equation}\label{dispersive1}
{\cal L}_{Disp} = \frac{\mu\, m_e^4}{4\pi\alpha} \,\,  \left [\partial_\alpha  h({\mathfrak F},{\mathfrak G})\right] \, \,   
F^{\alpha}_{\beta}F^{\beta\gamma}  \, \,
\left [\partial_\gamma   h({\mathfrak F},{\mathfrak G})\right].
\end{equation}
Here  $\mu$  is a coefficient that will be identified as 
\begin{equation}\label{mu}
\mu =\frac{4}{135 \pi}  \alpha
\end{equation}
by comparing  Eq.(\ref{1}) derived in terms of the invariant photon mass  with the corresponding result given by 
Eq.(\ref{KdV}) in Sec.\ref{SolCon}, and $h$  is a  function of the two Lorentz invariants 
${\mathfrak F}, \, {\mathfrak G}$. \, In the following we expand $h$ 
in a Taylor series and keep only  linear terms in  ${\mathfrak F}$ and  ${\mathfrak G}$.

\subsection{One dimensional, single polarization case} 
\label{1D}

In the case of an electromagnetic  configuration that depends on a single spatial coordinate, say $x$,  
and where the fields correspond to a single polarization state ${\mathfrak G}$ vanishes 
and Eq.(\ref{dispersive1}) becomes
\begin{equation}\label{dispersive2}
{\cal L}_{Disp||} =  -\frac{\mu\, m_e^4}{4\pi\alpha}\,  \left\{B^2 \left[\partial_x \left(E^2- B^2\right)\right]^2   
+ E^2 \left[\partial_t \left(E^2- B^2\right)\right]^2  
- 2 EB\left[\partial_x \left(E^2- B^2\right) \partial_t \left(E^2- B^2\right)\right] \right\},
\end{equation}
while 
\begin{equation}\label{dispersive2b}
{\mathcal{L}_{0}} +\mathcal{L}_{HE}=  -\frac{m_e^4}{4\pi\alpha}\, \left[\frac{B^2 - E^2}{2} 
 - \, \epsilon_2 \, \frac{(B^2 - E^2)^2}{4} + \, \epsilon_3 \, \frac{(B^2 - E^2)^3}{8}\right],
\end{equation}
where  
\begin{equation}\label{def1}
\epsilon_2=\frac{2 \alpha}{45 \pi} ,  \quad {\rm and} \quad
\epsilon_3 =\re{-}\frac{32 \alpha }{315 \pi}\ .
 \end{equation}

\section{Action functional  in light-cone variables}
\label{basic}

Here as in previous articles, see e.g. Ref. \cite{PRDS,HKAD2019,Hodo, Hodo-bis}, 
for a spatially one dimensional  configuration  it is convenient to introduce light cone variables defined as 
\begin{equation}
\label{def2}
x^+ = (x + t)/\sqrt{2},  \quad  x^- = (x - t)/\sqrt{2} \, ,\end{equation}
with corresponding derivatives
\begin{equation}
\label{def2a}
 \partial_x  = (\partial_+ + \partial_-)/\sqrt{2} \,  , 
 \quad
  \partial_t  = (\partial_+ - \partial_-)/\sqrt{2} \, ,
\end{equation}
where  $\partial_\pm = \partial /\partial x^\pm$.
For  a configuration corresponding to a single transverse polarization state  
we choose  the four vector potential to have only a component in the $z$ 
direction which we denote in normalized form by $a(x^+,x^-)$. 
Then the electric field $E$ is in the $z$ direction  while  the magnetic field $B$ is along $y$.  

We define the field variables 
\begin{equation}\label{def3}
w(x^+,x^-) = \partial _+ a(x^+,x^-), \quad  u(x^+,x^-) = \partial _- a(x^+,x^-),
\end{equation}
i.e. the electric and magnetic fields are
\begin{equation}\label{def3a}
 E = (u - w)/\sqrt{2} , \quad  B =  - (u + w)/\sqrt{2}.
\end{equation}
Note that by construction we recover Faraday's law in the form 
\begin{equation}\label{Faraday}
\partial _{-} w =  \partial _{-+}a = \partial_{+} u.
\end{equation}

By including both ${\mathcal{L}_{0}} +\mathcal{L}_{HE}$ and ${\cal L}_{Disp||}$,
 the electromagnetic action ${\cal A}$ expressed in the 
$x^+,\,  x^-, \,\,  w,\,  u\,\, $ variables takes the form
\begin{equation} \label{Action}
{\cal A}(a) = \frac{m^4}{4\pi\alpha}\iint_{\cal D} dx^+\, dx^- \mathcal{L}_T(a,a',a''),  
\end{equation}
where $a'$ stands symbolically for $w = \partial_+a $ and $u = \partial_-a$, 
while $a''$ stands for $\partial_+w = \partial_{++}a$, \, $\partial_- w= \partial_+u = \partial_{+ -}a$ \,  and  
$\partial_-u= \partial_{--}a$. 
After eliminating the common multiplicative factor in 
Eqs.(\ref{dispersive2},\ref{dispersive2b}) the Lagrangian  $\mathcal{L}_T$ reads 
\begin{equation}
\mathcal{L}_T(a,a',a'')=-uw+ \epsilon_2 (uw)^2- \epsilon_3 (uw)^3  
- \mu  \left\{w^2 \left[\partial_- (uw)\right]^2   +u^2 \left[\partial_+ (uw)\right]^2  + 2uw\left[\partial_+ (uw) \right]
\left[\partial_- (uw)\right] \right\}.
\label{Full-Lagrangian0}
\end{equation}
 
 The dependence of  Lagrangian  $\mathcal{L}_T$  on the first and on the second 
 order derivatives of the vector potential $a(x^+,x^-)$ can be made explicit  by rewriting (see Appendix \ref{A1})
$\mathcal{L}_T(a,a',a'')$ as 
\begin{align} &
 \mathcal{L}_T(a,a',a'')=-(\partial_+ a)(\partial_- a)+\epsilon_2 [(\partial_+ a)(\partial_- a)]^2-\epsilon_3 [(\partial_+ a)(\partial_- a)]^3 \label{Lagrangian-a} \\  
& - \mu  \left[ (\partial_+ a)^2 (\partial_{- -}a) +2(\partial_-a) (\partial_+ a)(\partial_{+ -} a)    + (\partial_-  a)^2(\partial_{+ +}a) \right]^2  .\nonumber 
\end{align}

Varying the action ${\cal A}(a)$  with respect to the vector potential $a$ and imposing that at the boundaries 
$\delta{\cal D}$ of the domain ${\cal D}$ under consideration both 
$\delta a$ and $\partial_+ \delta a,\, \partial_-  \delta a$ vanish, 
we obtain the wave equation for the vector potential 
\begin{equation}
- \partial_+ \frac{ \partial \, \mathcal{L}_T}{\partial \, (\partial_+a)}  - \partial_- \frac{ \partial \, \mathcal{L}_T}
{\partial \, (\partial_-a)} 
+ \partial_{+ +} \frac{ \partial \, \mathcal{L}_T}{\partial \, (\partial_{+ +}a)} +  \partial_{- -} 
\frac{ \partial \, \mathcal{L}_T}{\partial \, (\partial_{--}a)}\, + \partial_{+ -} \frac{ \partial \, \mathcal{L}_T}
{\partial \, ( \partial_{+ -}a)}= 0 ,
\label{Full-Lagrangian1}
\end{equation}
where  $\partial_{+}a = w,\, \, \partial_{-}a= u \, $ and $\, \partial_{+ +}a= \partial_+w,\, \,  
\partial_{+ -}=\partial_+u = \partial _- w, \, \,  \partial_{--}a= \partial_-u$ are treated as independent 
variables in the differentiation. The explicit form of the derivatives of $ {\mathcal{L}_T}$ in Eq.(\ref{Full-Lagrangian1}) 
are given in Appendix \ref{A2} 

\subsection{Field equations in the $u$, $w$ variables}\label{Feq}

The Lagrangian  $\mathcal{L}_T(w,u)$  in Eq.(\ref{Full-Lagrangian0}) can be rewritten as 
\begin{equation}  \label{Lagrangian-fields}
 \mathcal{L}_T(u,w)=-\left[uw -\epsilon_2 (uw)^2+\epsilon_3 (uw)^3  + 
 \mu  \left( w^2 \partial_- u+ uw\partial_+u + uw \partial_-w   + u^2\partial_+ w\right)^2 \right],
\end{equation}
and the wave equation for the vector potential (\ref{Full-Lagrangian1})  can be written in terms of 
 the  field variables $u$ and $w$   as  
\begin{align} 
\label{full wave} 
& \partial_{+}
\left\{u\left[1  - 2 \epsilon_2 uw + 3\epsilon_3 (uw)^2\right]\right\} 
 + \partial_{-}\left\{w\left[1  - 2 \epsilon_2 uw  + 3 \epsilon_3 (uw)^2 \right]\right\}  \\&
 + 4\mu  \left\{\partial_+[w(\partial_- (uw))^2 +  u \partial_+(uw) \partial_-(uw)]  +   
 \partial_-[u(\partial_+(uw))^2  + w \partial_+(uw) \partial_-(uw)]  \right\}\nonumber \\  
&-2\mu\left\{ \partial_{++}[  u^3  \partial_+ (uw) + u^2 w \partial_- (uw)] +
\partial_{--}[w^3  \partial_- (uw) +   u w^2 \partial_+ (uw) ]  +2 \partial_{+-}[uw^2  \partial_- (uw) + u^2w \partial_+ (uw)]\right\} ,\nonumber
\end{align}
which can be derived by rearranging Eq.({\ref{explicit1})  in Appendix  \ref{A2}.

\section{Scattering solutions for counter-streaming finite length pulses}\label{scatter}

The asymptotic effect  of the interaction between two counter-propagating 
electromagnetic pulses  with a finite length con be derived directly from 
Eq.(\ref{Full-Lagrangian1})
assuming  that for large $|x_\pm |$ there is  no superposition between the pulses so that  the vector potential 
$a(x^+,x^-)$ can be written as $a(x^+,x^-) =  a_+(x^+)  + a_-(x^-)$. 
Integrating Eq.(\ref{Full-Lagrangian1}) over $x^+$ we obtain
\begin{equation}
\frac{\partial \, \mathcal{L}_T}{\partial \, (\partial_+a)} \Bigr\rvert^{+\infty}_{-\infty} = 
- \partial_- \int^{+\infty}_{-\infty}\, d x^+ 
\frac{ \partial \, \mathcal{L}_T}{\partial \, (\partial_-a)} 
+  \partial_{- -}  \int^{+\infty}_{-\infty}\, d x^+ 
\frac{ \partial\, \mathcal{L}_T}{\partial \, (\partial_{--}a)} .
\label{Full-Lagrangian2}
\end{equation}
Then, considering a  perturbative   expansion around $a_0(x^+,x^-) =  a_{0+}(x^+)  + a_{0-}(x^-)  $,  
and using Eq.(\ref{explicit}),   we find 
\begin{align}  & \label{Full-Lagr3}
 \partial_-a \Bigr\rvert^{+\infty}_{-\infty} =  
 \partial_-  \int^{+\infty}_{-\infty}\, d x^+ \left [ \epsilon_2 (\partial_+a_0)^2(\partial_-a_0 ) 
 -\frac{3}{2} \epsilon_3 (\partial_+a_0)^3(\partial_-a_0)^2 
- \mu  (\partial_-  a_0)^3(\partial_{+ +}a_0)^2\right ]   \\&
+ 2\mu   \partial_{- -}  \int^{+\infty}_{-\infty}\, d x^{+}  (\partial_+a_0)^4 (\partial_{- -} a_0)  \nonumber. 
\end{align}
The terms on the r.hs. arise from the Heisenberg-Euler  Lagrangian  (the $\epsilon_2,\epsilon_3$  terms)  
and from the dispersive additional contribution   
(the $\mu$  terms) given in Eq.(\ref{dispersive2}). In deriving Eq.(\ref{Full-Lagr3}) we used the fact  that 
$\partial_{+-} a_0 = 0$ and that
 two terms proportional to $(\partial_+ a_0)^2 (\partial_{+ +}a_0) $ 
are  total derivatives  with respect to $x^+$  and thus  do not contribute to the $x^+$ 
integral when asymptotically there is no superposition 
between the pulses. The term on the l.h.s. represents  the lowest order change 
of the vector potential  pulse propagating along the positive $x$ 
direction due to the scattering with the counter-propagating pulse, and the factor 2 in front 
of it  arises from the contribution of the classical 
electromagnetic Lagrangian. Expressed in terms of the  electromagnetic field variables 
Eq.(\ref{Full-Lagr3}) reads
\begin{align} & \label{Full-Lagr3bis}
 u \Bigr\rvert^{+\infty}_{-\infty} =   \epsilon_2 (\partial_- u_0) \int^{+\infty}_{-\infty}\, d x^+  w_0^2
 -\frac{3  }{2} \epsilon_3(\partial_- u_0^2) \int^{+\infty}_{-\infty}\, d x^+  w_0^3 
 -  \mu  ( \partial_-u_0^3)
 \int^{+\infty}_{-\infty}\, d x^+ (\partial_+ w_0)^2    \\&  
+\mu  ( \partial_{- --} u_0) \, \int^{+\infty}_{-\infty}\, d x^+   w_0^4   . \nonumber
\end{align}
where $w_0 = w_0(x^+) = \partial_+ a_{0}(x^+)$ and $u_0 = u_0(x^-) = \partial_- a_{0}(x^-)$.  
A corresponding equation can be derived for $ w \rvert^{+\infty}_{-\infty} $.  
The first term on the r.h.s. of Eq.(\ref{Full-Lagr3bis}) corresponds to the  standard phase shift due 
to reduced propagation velocity during the interaction phase \cite{Hodo,Shukla,Mark}
 while the second, if  the integral of $w_0^3$ does not vanish,  to the six-photon interaction  
 contribution to the harmonic generation mechanism discussed e.g. in Ref.\cite{harm}.  
The third terms corresponds to a new harmonic generation process  that depends on the square of the 
derivative of the field amplitude  of the counter-propagating pulse, while the fourth term provides 
a dispersion correction to the phase shift given by the first term and corresponds to a widening of the pulse.

\section{Solutions in constant cross fields}
\label{SolCon}

The system of Eqs.(\ref{Faraday},\ref{full wave}) admits solutions in the form of the progressive nonlinear waves that
 propagate  with constant ``velocity''  $S>0$, i.e. with  functions $u$ and $w$ that depend  on the variable
\begin{equation}
\label{phase-psi}
\psi=x^{-}+S x^{+}=\frac{1}{\sqrt{2}}\left[x(1+S)-t(1-S) \right].
\end{equation}
In the $x$-$t$ variables the wave propagates with the velocity equal to $(1-S)/(1+S)$.

Using Eq.(\ref{Faraday}) we obtain a relationship between $u(\psi)$ and $w(\psi)$
\begin{equation}
\label{uw-psi}
w=S u +W_0,
\end{equation}
where $W_0$ is   constant  which corresponds to a constant 
 cross field configuration with equal amplitude electric and magnetic fields,  
 $E_0=B_0=\sqrt{2} W_0$. The Poynting vector  $c {\bf E} \times {\bf B}/4\pi$ of the cross  
 field   configuration that is taken to model a low frequency wave  is directed in the negative
direction along the $x$-axis so that 
${\bf E}={\bf e}_z E_0$, ${\bf B}={\bf e}_y B_0$, where ${\bf e}_y$ and ${\bf e}_z$  
are unit vectors in the $y$ and $z$ directions. The high frequency electromagnetic wave 
described by the variables $w$ and $u$ in Eq.(\ref{uw-psi}) 
 propagates in the positive direction along the $x$-axis.

Assuming for the sake of simplicity that the amplitude of the high frequency wave amplitude is much smaller  than 
the cross  field amplitude ($|u|,|w| \ll W_0$)  we obtain from Eqs.(\ref{full wave},\ref{uw-psi})
\begin{equation} 
\label{KdV} 
 2 \mu W_0^4\partial_{---}u=
\partial_{+}\left(u  - 2 \epsilon_2 W_0 u^2 \right)
 + \partial_{-}\left(S u  -  2 \epsilon_2 W_0^2 u + 3\epsilon_3 W_0^3 u^2 \right).
\end{equation}
Using the ansatz (\ref{phase-psi}) we obtain
\begin{equation} 
\label{KdV-wave} 
  \mu W_0^4 u'''=
\left[S-\epsilon_2 W_0^2  - \left( 2 S  \epsilon_2 W_0-3 \epsilon_3 W_0^3 \right)u\right] u',
\end{equation}
where a prime stands for differentiation with respect to $\psi$. This is the well known Korteveg-de Vries equation 
for the stationary nonlinear wave propagating with constant velocity 
$S$ (see Refs. \cite{KdV1885, WHIT1974, SPN1984} 
and Ref. \cite{PRDS} for the case of the KdV solitons in the QED vacuum).
Integration of Eq.(\ref{KdV-wave}) over $\psi $ yields 
\begin{equation} 
\label{KdV-wave-INT1} 
  \mu W_0^4 u''=
\left(S-\epsilon_2 W_0^2\right) u  - \left( S  \epsilon_2 W_0-3 \epsilon_3 W_0^3/2 \right)u^2+C_1.
\end{equation}
Multiplying this equation on $u'$ and integrating over $\psi $ yields
\begin{equation} 
\label{KdV-wave-INT2} 
  \mu W_0^4 (u')^2=
\left(S-\epsilon_2 W_0^2\right) u^2  - \left( 2S  \epsilon_2 W_0/3-  \epsilon_3 W_0^3 \right)u^3+2 C_1 u+C_2,
\end{equation}
where $C_1$ and $C_2$ are constants. 

Choosing $C_1 = C_2 =0$ we find the solution of Eq.(\ref{KdV-wave-INT2}) in the form of a KdV soliton. It reads
\begin{equation} 
\label{KdV-soliton} 
  u(x^++Sx^-)=\frac{3 (S-\epsilon_2 W_0^2)}{(2 \epsilon_2 W_0-3 \epsilon_3 W_0^3)}\cosh^{-2}\left[\frac{\sqrt{S-\epsilon_2 W_0^2}(x^++Sx^-)}{2\sqrt{\mu W_0^4}} \right] ,
\end{equation}
with $ W_0^2 < (2/3)\, (\epsilon_2/\epsilon_3)$.
The soliton amplitude $u_0$ and width $l_0$ are given by 
\begin{equation} 
\label{KdV-soliton-u0l0} 
 u_0=\frac{3 (S-\epsilon_2 W_0^2)}{(2 \epsilon_2 W_0-3 \epsilon_3 W_0^3)}\qquad {\rm and } \qquad l_0=2\sqrt{\frac{\mu W_0^4}{S-\epsilon_2 W_0^2}}.  
\end{equation}
We see that the parameter $S$ determining the soliton propagation velocity depends on the soliton 
amplitude $u_0$ as 
\begin{equation} 
\label{KdV-soliton-S} 
 S=\epsilon_2 W_0^2+\left(\frac{2}{3}\epsilon_2 W_0-\epsilon_3 W_0^3 \right)u_0 .
 \end{equation}
 The soliton propagation velocity depends on $W_0$ and $u_0$ as 
\begin{equation} 
\label{KdV-soliton-V} 
V=\frac{1-S}{1+S} =\frac
{1-\epsilon_2 W_0^2-(2 \epsilon_2 /3-\epsilon_3 W_0^2) W_0 u_0}
{1+\epsilon_2 W_0^2+(2 \epsilon_2 /3-\epsilon_3 W_0^2) W_0 u_0}
\approx
1-2\epsilon_2 W_0^2-\frac{4}{3}\epsilon_2 W_0 u_0.
 \end{equation}
Substituting $S$ from Eq.(\ref{KdV-soliton-S}) to the expression for $l_0$ given by Eq.(\ref{KdV-soliton-u0l0}) we 
find the soliton width. It reads
\begin{equation} 
\label{KdV-soliton-l0} 
l_0=\sqrt{\frac{4\mu W_0^3}{(2 \epsilon_2 -3 \epsilon_3 W_0^2) u_0}} ,
 \end{equation}
i. e. $l_0\approx \lambdabar_C(W_0^3/u_0)^{1/2}$. In other words a typical energy of the photons constituting 
the soliton is approximately equal to $\hbar \omega_{\gamma}\approx m_e c^2(W_0^3/u_0)^{1/2}$.

\section{Conclusions}\label{concl}
We use  a  Lagrangian  that involves higher order derivatives of the wave vector potential and that is constructed 
so as to include the quantum nonlinearity parameter dependency of the invariant photon mass. 
This Lagrangian  allows 
us to describe  dispersive effects  in the interaction of two counter-propagating  light pulses by a nonlocal  
extension of the nonlinear wave equation that is derived from the Heisenberg-Euler Lagrangian.  
In addition, in the case of a finite amplitude wave impinging on  large cross fields, we   
show that Korteveg-de Vries soliton solutions  can be consistently derived from these field equations by considering 
a proper ordering of the amplitude of the impinging wave and of its space-time 
coordinate dependence in terms of the amplitude of the cross fields.  

An extension of this procedure so as to include higher order derivatives and higher powers of the fields amplitude 
 than those considered in this article  could  be of interest when searching 
for novel light soliton solutions, such as e.g ``compactons'',  i.e. solitons with finite wavelength \cite{compact}.
Such an extension could be written in the formal way
\begin{equation}\label{extension}
{\cal L}_{HE/D} =  
\frac{m^4}{4\pi\alpha}\epsilon_2\, h\left(F_{\mu\nu}F^{\mu\nu}, \epsilon^{\mu \nu \kappa \lambda}F_{\mu\nu}F_{\kappa \lambda} \right)\,
\left[ 1 + f (\,\overleftarrow{ \partial_\alpha}  F^{\alpha}_{\beta}F^{\beta\gamma}\overrightarrow{ \partial_\gamma} \,)
\right] \,
h\left(F_{\mu\nu}F^{\mu\nu} , \epsilon^{\mu \nu \kappa \lambda}F_{\mu\nu}F_{\kappa \lambda} \right) ,
\end{equation}
where the function $h$ is related to the usual Heisenberg-Euler  asymptotic expansion while $f$ is a ``function'' 
of the differential operator 
$\overleftarrow{ \partial_\alpha}  F^{\alpha}_{\beta}F^{\beta\gamma}\overrightarrow{ \partial_\gamma}$ 
(related to the relativistic  $\chi$ invariant) where the arrows indicate left or right action. 
The function $f$ should be related 
to the expansion of the invariant  photon mass   in Eq.(\ref{mass}), see Ref.\cite{Ritus}. 
The Lagrangian ${\cal L}_{HE/D} $ 
is gauge invariant and is Lorentz invariant. In addition its contribution to the wave equation vanishes in the case 
of a plane wave in which case 
$ F_{\mu\nu}F^{\mu\nu} = \varepsilon^{\mu \nu \kappa \lambda}F_{\mu\nu}F_{\kappa \lambda}   =0$.


\begin{acknowledgments}

The work is supported by the project High Field Initiative
(CZ.02.1.01/0.0/0.0/15\_003/0000449)
from the European Regional Development Fund.
\end{acknowledgments}

\appendix
\section{Explicit form of the Lagrangian ${\cal L}_{TS}$}\label{A1}

We rewrite the Lagrangian    $\mathcal{L}_T$  in   Eq.(\ref{Full-Lagrangian0}) explicitly as a function of 
$\partial_+a, \, \partial_-a,\, \partial_{++}a, \,  \partial_{+-}a,\,  \partial_{--}a$
\medskip
\begin{align}  & 
 \mathcal{L}_T(a,a',a'')=-\left\{(\partial_+ a)(\partial_- a)-
 \epsilon_2 \left[(\partial_+ a)(\partial_- a)\right]^2-\epsilon_3 \left[(\partial_+ a)(\partial_- a)\right]^3  \right. \label{bis}\\  
&\left. + \mu  \left[(\partial_+ a)^2 ( (\partial_{--}a)(\partial_+ a) +(\partial_-a)(\partial_{+-} a)  )^2   
+(\partial_- a)^2 ((\partial_{-+} a)(\partial_+ a) + (\partial_-  a)(\partial_{++}a) )^2 \right.  \right. \nonumber \\  
&\left. \left. + 2(\partial_- a)(\partial_+ a) ( (\partial_{--}a)(\partial_+ a) +(\partial_-a)(\partial_{+-} a)  )
((\partial_{-+} a)(\partial_+ a) + (\partial_-  a)(\partial_{++}a) )
 \right] \right\} ,\nonumber
\end{align}
which can be rewritten as
\begin{align} &
 \mathcal{L}_T(a,a',a'')=-\left\{(\partial_+ a)(\partial_- a)-\epsilon_2 ((\partial_+ a)(\partial_- a))^2
 +\epsilon_3 ((\partial_+ a)(\partial_- a))^3  \right. \label{Lagrangian-abis} \\  
&\left. + \mu  \left[ ((\partial_+ a)^2 (\partial_{--}a) +2(\partial_-a) (\partial_+ a)(\partial_{+-} a)    
+ (\partial_-  a)^2(\partial_{++}a) )^2  \nonumber 
 \right] \right\} ,
\end{align}
and  can be re-expressed in terms of the electromagnetic fields as
\begin{equation}  \label{Lagrangian-fieldsb}
 \mathcal{L}_T(w,u)=-\left\{uw -\epsilon_2 (uw)^2+\epsilon_3 (uw)^3  + \mu  \left[ w^2 \partial_- u+ uw\partial_+u + uw \partial_-w   + u^2\partial_+ w\right]^2 \right\}.
\end{equation}

\section {Derivation of the field equations} \label{A2}
From Eq.(\ref{Lagrangian-a}) we find
\begin{align}\label{explicit}
& \frac{ \partial \, \mathcal{L}_T}{\partial \, (\partial_+a)} = -(\partial_-a)\left[1  - 2 \epsilon_2( \partial_+a)(\partial_-a)  + 3\epsilon_3 (\partial_{+}a)^2(\partial_{-}a)^2 \right] \\&\qquad
- 4\mu \left\{\left[(\partial_{+} a)(\partial_{--} a)    + (\partial_-  a)(\partial_{+-}a)\right]
\left[(\partial_+ a)^2 (\partial_{--}a) +2(\partial_-a) (\partial_+ a)(\partial_{+-} a)    + (\partial_-  a)^2(\partial_{++}a) \right]\right\},  \nonumber \\&  ~\nonumber \\
&  
 \frac{ \partial \, \mathcal{L}_T}{\partial \, (\partial_-a)} = - \left[\partial_+a)(1  - 2 \epsilon_2 (\partial_+a)(\partial_-a ) +3\epsilon_3 (\partial_+a)^2(\partial_-a)^2 \right]\nonumber  \\
 &  \qquad
-4 \mu \left\{
\left[(\partial_+ a)(\partial_{+ -} a)    + (\partial_-  a)(\partial_{+ +}a)\right]\left[(\partial_+ a)^2 (\partial_{- -}a) 
+2(\partial_-a) (\partial_+ a)(\partial_{+ -} a)    + (\partial_-  a)^2(\partial_{+ +}a) \right]\right\} ,\nonumber
\\&~\nonumber \\
&
\frac{ \partial \, \mathcal{L}_T}{\partial \, (\partial_{+ +}a)} 
 =  -2\mu 
 \left\{(\partial_-a)^2 \left[(\partial_+ a)^2 (\partial_{- -}a) +2(\partial_-a) (\partial_+ a)(\partial_{+ -} a)    
 + (\partial_-  a)^2(\partial_{+ +}a) \right]\right\} , \nonumber \\& ~\nonumber \\
 &
\frac{ \partial \, \mathcal{L}_T}{\partial \, (\partial_{- -}a)} =
- 2 \mu
\left\{(\partial_+a)^2 \left[(\partial_+ a)^2 (\partial_{- -}a) +2(\partial_-a) (\partial_+ a)(\partial_{+ -} a)    
+ (\partial_-  a)^2(\partial_{+ +}a) \right]\right\}  , 
\nonumber  \\& ~\nonumber \\
&
\frac{ \partial \, \mathcal{L}_T}{\partial \, (\partial_{+ -}a)} =
 -4\mu 
    \left\{(\partial_+a)(\partial_-a) \left[(\partial_+ a)^2 (\partial_{- -}a) +2(\partial_-a) (\partial_+ a)(\partial_{+ -} a)    
    + (\partial_-  a)^2(\partial_{+ +}a) \right]\right\} . \nonumber
\end{align}
\medskip

 Using Eq.(\ref{explicit}),  reintroducing the field variables $u,w$,  and  defining for the sake 
 of notational compactness 
 \begin{equation}
 {\cal M}(u,w) = w^2 \partial_- u + uw \partial_- w +  uw \partial_+ u  + u^2 \partial_+w=  
 w \partial_-(uw) + u \partial_+(uw),
 \end{equation}
 from Eq.(\ref{Full-Lagrangian1}) we find
\begin{align}\label{explicit1}& ~\quad 
\partial_{+}\left\{u\left[1  - 2 \epsilon_2 wu  + 3\epsilon_3 (wu)^2 \right]\right\}  
+ \partial_{-}\left\{w\left[1  - 2 \epsilon_2 wu  + 3\epsilon_3 (wu)^2 \right]\right\}  \\& 
+2\mu \left\{2\partial_{+}\left[ (w\partial_- u + u \partial_- w) {\cal M}(u,w)\right]  
+2 \partial_-[ (w\partial_+ u + u \partial_+ w) {\cal M}(u,w)] \right.  \nonumber \\& 
-\left. \partial_{+}\partial_{+} [u^2 {\cal M}(u,w)]  -  \partial_{-}\partial_{-} [w^2M(u,w)]  - 
2  \partial_+\partial_- [uw {\cal M}(u,w)]\right\} . \qquad  \qquad \nonumber
\end{align}

\section{Ghosts branches}
\label{GHOSTS}

In the following two sections we discuss ghost branches of the Higher order Lagrangians considered in this article.
Implementing  the high derivative term (\ref{L:MME}) of  the Lagrangian (\ref{Lag2})  in the Heisenberg-Euler Lagrangian leads to 
\begin{equation}  
\label{Lagrangian-MME}
 \mathcal{L}(u,w)=
 -uw +\epsilon_2 (uw)^2-\epsilon_3 (uw)^3  
 -\tilde{\mu}
 \left[ (\partial_{+}u)(\partial_{-}w)+u\partial_{+-}w+w\partial_{+-}u\right],
\end{equation}
where $\tilde{\mu}={M^{-2}}$ is proportional to the fine structure constant $\alpha$., leading to the system 
of wave equations
\begin{equation}\label{Faraday1}
\partial _{-} w = \partial_{+} u,
\end{equation}
\begin{equation}
 \partial_{+}
\left\{u\left[1  - 2 \epsilon_2 uw  + 3\epsilon_3 (uw)^2\right]\right\} 
 + \partial_{-}\left\{w\left[1  - 2 \epsilon_2 uw  + 3\epsilon_3 (uw)^2 \right]\right\}+4 \tilde{\mu} \partial_{++-} u.
\end{equation}
 
  We consider an  electromagnetic wave counter-propagating  with respect to a large  amplitude cross field 
low frequency electromagnetic wave, i.e. we assume that 
\begin{equation}\label{W0-APPF}
a(x^+, x^-)  = W_0 x^+ + {\tilde a} (x^+, x^-) . 
\end{equation}
Within the linear wave approximation, which requires $\{|\partial_-{\tilde a}|\,,\,|\partial_+{\tilde a}|\}\ll W_0$,  
Eqs. (\ref{Faraday1},\ref{W0-APPF}) can be reduced to the equation
\begin{equation}\label{F:wave-lin}
\partial_{+} {\tilde a} - \epsilon_2 W_0^2 \partial_{- } {\tilde a} +4 \tilde{\mu} \partial_{++-} {\tilde a}
 = 0 ,
 \end{equation}
which can be rewritten in  $x,t$ variables as 
\begin{equation}\label{F:wave-linxt}
(1+\epsilon_2 W_0^2)\partial_{t} {\tilde a} +(1-\epsilon_2 W_0^2)\partial_{x} {\tilde a} +2 \tilde{\mu} 
(\partial_{t}+\partial_{x}) (\partial_{tt}-\partial_{xx}){\tilde a}
 = 0 .
 \end{equation}
 The corresponding dispersion equation giving a relationship between the wave frequency $\omega$ and wave
  number $k$, i.e. ${\tilde a}(x,t) \propto \exp(-i(\omega t- k x))$, has a form
\begin{equation}
\label{F:dispeq-linxt}
(1+\epsilon_2 W_0^2)\omega -(1-\epsilon_2 W_0^2)k -2 \tilde{\mu} \left(\omega-k)(\omega^2-k^2\right)
 = 0 .
 \end{equation}
 It is convenient to rewrite this equation in terms of $\Omega=(\omega+k)/\sqrt{2}$ and 
 $Q=(\omega-k)/\sqrt{2}$. In this case, we have  ${\tilde a}(x^+,x^-) \propto \exp(-i(\Omega x^{+}+Q x^{-}))$. 
 The dispersion equation can be written as
\begin{equation}
\label{F:wave-OmQu}
\Omega - \epsilon_2 W_0^2 Q +4 \tilde{\mu} \Omega^2 Q,
 = 0 ,
 \end{equation}
 whose solution gives for two branches 
\begin{equation}\label{F:dequ-OmQu}
\Omega_{\pm}=\frac{-1 \pm \sqrt{1+16 \epsilon_2 W_0^2}}{8 \tilde{\mu} Q}.
 \end{equation}
  In the long-wavelength limit $Q\to 0$ and/or in the limit of week dispersion $\tilde{\mu}\to 0$ 
 the frequency $\Omega_+$ equals
\begin{equation}
\label{F:dequ-Om-p}
\Omega_{+}=\epsilon_2 W_0^2 Q- 4  \tilde{\mu} \epsilon_2 W_0^4 Q^3+...\,,
  \end{equation}
 while the frequency $\Omega_-$ corresponds to the ghost branch, 
\begin{equation}
\label{F:dequ-Om-m}
\Omega_{-}=\frac{1}{4 \tilde{\mu} Q}+\epsilon_2 W_0^2 Q\, ...\,.
  \end{equation}
 In terms of $\omega_-$ and $k_-$ this results in the dispersion equation 
\begin{equation}
\label{F:dequ-Om-m}
\omega_{-}=\sqrt{k_-^2+\frac{1}{4 \tilde{\mu}}},
  \end{equation}
 i.e. it describes the photons with the ``mass'' which tends to infinity when $\tilde{\mu}  \to 0$. Using Eq.(\ref{mu}) 
 we  find that the electromagnetic field with the photon energy corresponding to  the ``mass'', 
 $\hbar \omega_-=\sqrt{135 \pi/4 \alpha}m_ec^2$ which cannot be described within the Heisenberg-Euler  Lagrangian 
 paradigm in contrast to  the the wave  corresponding to the branch given by Eq.(\ref{F:dequ-Om-p}).

\section{Lorentz invariant solutions}\label{A3}

Following a procedure adopted in Ref.\cite{Hodo} we may look  for self similar solutions where the vector potential 
$a(x^+,x^-)$ depends only  on the combination  $x^+x^-= \rho$ which is invariant under Lorentz boosts along $x$.

Using the following formulae for Lorentz invariant solutions with 
$a(x^+,x^-) = {\hat a}(\rho)$ and ${\hat a}^\prime = d {\hat a}/d\rho$
\begin{align}\label{form1}
& \partial_+ a = x^- {\hat a}^\prime,    \quad \partial_- a = x^+\, {\hat a}^\prime,  \quad  \rho ({\hat a}^\prime)^2  = uw,\quad \partial_{++}  a 
=( x^-)^2\, {\hat a}^{\prime \prime}, \quad  \partial_{--} a =( x^+)^2\, {\hat a}^{\prime \prime}, \quad  \partial_{+-}  a =  {\hat a}^\prime+ \rho  
{\hat a}^{\prime \prime}
\nonumber\\
&  (\partial_+ a)^2 (\partial_{--}a) +2(\partial_-a) (\partial_+ a)(\partial_{+-} a)    + (\partial_-  a)^2(\partial_{++}a) =
4 \rho^2  ({\hat a}^\prime)^2{\hat a}^{\prime \prime}  + 2 \rho ({\hat a}^\prime)^3 ,
\end{align}
we obtain
\begin{align}
\label{form2}
& \frac{ \partial \, \mathcal{L}_T}{\partial \, (\partial_+a)} = 
- x^+\left[{\hat a}^\prime - 2 \epsilon_2\rho ({\hat a}^\prime)^3 + 3 \epsilon_3\rho^2 ({\hat a}^\prime)^5  
+ 4 \mu ( ({\hat a}^\prime)^2 + 2\rho  {\hat a}^\prime {\hat a}^{\prime\prime})(4 \rho^2  ({\hat a}^\prime)^2{\hat a}^{\prime \prime}  + 2 \rho ({\hat a}^\prime)^3)\right]
  \\&
=\, - x^+\left[{\hat a}^\prime - 2 \epsilon_2\rho ({\hat a}^\prime)^3 + 3 \epsilon_3\rho^2 ({\hat a}^\prime)^5\right] 
-  8 \mu  x^+ \rho  {\hat a}^\prime \left[ (\rho (a^\prime)^2)^\prime\right]^2 
 \nonumber \\& 
 \frac{ \partial \, \mathcal{L}_T}{\partial \, (\partial_-a)} =
- x^-\left[{\hat a}^\prime - 2 \epsilon_2\rho ({\hat a}^\prime)^3 - 3 \epsilon_3\rho^2 ({\hat a}^\prime)^5\right] 
-  8 \mu  x^- \rho  {\hat a}^\prime \left[ (\rho (a^\prime)^2)^\prime\right]^2 \nonumber\\&
\frac{ \partial \, \mathcal{L}_T}{\partial \, (\partial_{+ +}a)} 
 =  -2\mu  (x^+)^2 ({\hat a}^\prime)^2(4 \rho^2  ({\hat a}^\prime)^2{\hat a}^{\prime \prime}  
 + 2 \rho ({\hat a}^\prime)^3 ) = -4\mu \rho  (x^+)^2 
  ({\hat a}^\prime)^3(\rho (a^\prime)^2)^\prime, \nonumber \\&
\frac{ \partial \, \mathcal{L}_T}{\partial \, (\partial_-\partial_-a)} =-4\mu \rho  (x^-)^2
  ({\hat a}^\prime)^3(\rho (a^\prime)^2)^\prime, \nonumber \\& 
  \frac{ \partial \, \mathcal{L}_T}{\partial \, (\partial_+\partial_-a)} =  - 8\mu \rho^2 ({\hat a}^\prime)^3(\rho (a^\prime)^2)^\prime. \nonumber
\end{align}
Then from Eqs.(\ref{Full-Lagrangian1},\ref{form2}) we have  
\begin{equation}\label{form3} \left[\rho {\hat a}^\prime\right]^\prime = \left[\rho
{\cal H} (\rho, {\hat a}^\prime)\right]^\prime  -2\mu [{\cal K} (\rho,  {\hat a}^\prime,  {\hat a}^{\prime\prime})]
^\prime \end{equation}
where 
\begin{align}
\label{form4}
& {\cal H} (\rho,{\hat a}^\prime)   =  
2 \epsilon_2\rho ({\hat a}^\prime)^3 -3 \epsilon_3\rho^2 ({\hat a}^\prime)^5  
-8 \mu\rho  {\hat a}^\prime \left[ ({\hat a}^\prime)^2 + 2\rho  {\hat a}^\prime {\hat a}^{\prime\prime}\right]^2, \\&
 {\cal K} (\rho, {\hat a}^\prime,{\hat a}^{\prime\prime})   = 
4 \rho {\hat a}^\prime[  (\rho ({\hat a}^\prime)^2)^\prime]^2 + [\rho^2  {\hat a}^\prime [\rho^2  ({\hat a}^\prime)^4]^\prime ]^{\prime}  + 
\rho[\rho {\hat a}^\prime [\rho^2  ({\hat a}^\prime)^4]^\prime ]^\prime.\nonumber 
\end{align} 
A logarithmic-type solution   is obtained in a perturbative approach (see Ref.\cite{Hodo}) where, to zero order we have 
\begin{equation}
\label{form5} 
\left[\rho {\hat a}^\prime\right]^\prime =0,\quad 
{\rm which\, leads\, to} \quad {\hat a}(\rho) = C_1 + C_2 \ln(|\rho |) 
\end{equation}
while the higher order contribution on the r.h.s. of Eq.(\ref{form3})  are properly included  
by a renormalization procedure that leads to a modification of the argument of the logarithmnof the form 
${\hat a} = \ln{|\rho +g(\rho)|} $.


\begin{thebibliography}{99}

\bibitem{BLP-QED} V. B. Berestetskii, E. M. Lifshitz, and  L. P. Pitaevskii, {\it Quantum Electrodynamics} (Pergamon, New York, 1982).

\bibitem{VLG} V. L. Ginzburg, {\it  Theoretical physics and astrophysics}  (Pergamon Press, New York, 1979).

\bibitem{HeisenbergEuler} W. Heisenberg and H. Euler, {\it Zeit. f\"ur Phys.}, {\bf 98},  714 (1936).

\bibitem{Karplus1951} R. Karplus and M. Neumann, {\it  Phys. Rev.},  {\bf 83}, 776 (1951).

\bibitem{Schw51} J. Schwinger, {\it Phys. Rev.}, {\bf 82}, 664 (1951).

\bibitem{McK1963} J. McKenna and P. M. Platzman, {\it Phys. Rev.}, {\bf 129}, 2354 (1963). 

\bibitem{Erber1966} T. Erber, {\it Rev. Mod. Phys.}, {\bf 38}, 626 (1966).

\bibitem{Mamaev} S. G. Mamaev, V. M. Mostepanenko, and M. I. Eides,
{\it Sov. J. Nucl. Phys.},  {\bf 33}, 569 (1981).

\bibitem{Grib1980} A. A. Grib, S. G. Mamaev, and V. M. Mostepanenko, 
{\it Quantum Effects in Intense External Fields} (Moscow: Atomizdat, 1980).

\bibitem{Batt2018} R. Battesti, J. Beard, S. Boeser, N. Bruyant, and D. Budker, {\it et al.},
{\it Phys. Rep.}, {\bf 765}, 1 (2018).

\bibitem{Bernard2000} 
D. Bernard, F. Moulin, F. Amiranoff, A. Braun, J. P. Chambaret, G. Darpentigny, G. Grillon, S. Ranc, and F. Perrone, 
 {\it Eur. Phys. J. D}, {\bf 10}, 141 (2000).

\bibitem{CDanson2015} C. N. Danson, D. Hillier, N. Hopps, and D. Neely, 
{\it High Power Laser Sci. Eng.}, {\bf 3}, e3 (2015).

\bibitem{CDanson2019} C. N. Danson, et al.,
{\it  High Power Laser Sci. Eng.}, {\bf 7}, e54 (2019).
 
\bibitem{Heinzl2006} T. Heinzl, B. Liesfeld,  K.-U. Amthor, H. Schwoerer, R. Sauerbrey,  and A. Wipf,
{\it Optics Communications}, {\bf 267}, 318 (2006).

\bibitem{Schlenvoigt2016} H.-P. Schlenvoigt, T. Heinzl, U. Schramm, T. E. Cowan,  and R. Sauerbrey, 
{\it Phys. Scr.}, {\bf 91},  023010  (2016).

\bibitem{King} B. King and  T. Heintzl, {\it High Power Laser Science and Engineering}, {\bf 4}, e5 (2016).

\bibitem {Mourou} G. A. Mourou, T. Tajima,  and S. V. Bulanov, {\it Rev. Mod. Phys.}, {\bf 78},  309 (2006).

\bibitem{Marklund} M. Marklund and  P. K. Shukla, {\it Rev. Mod. Phys.}, {\bf 78}, 591 (2006).

\bibitem{DTom} D. Tommasini, A. Ferrando, and M. Seco, {\it Phys. Rev. A}, {\bf 77}, 042101 (2008).

\bibitem{Pare}  A. Paredes, D. Novoa, and D. Tommasini, {\it Phys. Rev. A}, {\bf 90}, 063803 (2014).

\bibitem{DiPiazzaReview}  A. Di Piazza, C. M\"uller, K. Z. Hatsagortsyan,  C. H. Keitel, 
{\it Rev. Mod. Phys.}, {\bf 84}, 1177 (2012).

\bibitem{BattRizz}  R. Battesti and C. Rizzo, {\it Rep. Prog. Phys.}, {\bf 76}, 016401 (2013).

\bibitem{Koga} J. K. Koga, S. V. Bulanov, T. Zh. Esirkepov, A. S. Pirozhkov, M. Kando, and N. N. Rosanov, 
{\it Phys. Rev. A}, {\bf 86}, 053823 (2012).

\bibitem{KarbsteinShai1} F. Karbstein and R. Shaisultanov, {\it Phys. Rev. D}, {\bf 91}, 113002 (2015).

\bibitem{KarbsteinShai2}  H. Gies, F. Karbstein, C. Kohlfuerst,  and N. Seegert, {\it Phys. Rev. D}, {\bf 97}, 076002 (2018).

\bibitem{Angioi2019} A. Angioi and A. Di Piazza, {\it Rend. Fis. Acc. Lincei}, {\bf 30}, 17 (2019).

\bibitem{SVBRFAL2019} S. V. Bulanov,  {\it Rend. Fis. Acc. Lincei}, {\bf 30}, 5 (2019).

\bibitem{ZHANG} P. Zhang, S. S. Bulanov, D. Seipt, A. V. Arefiev, and A. G. R. Thomas, {\it Phys. Plasmas} {\bf 27}, 050601 (2020).

\bibitem{ADP2018} A. Di Piazza, M. Tamburini, S. Meuren, and C. H. Keitel, {\it Phys. Rev. D}, {\bf 97}, 056028 (2018).

\bibitem{FK2015} F. Karbstein, H. Gies, M. Reuter, and M. Zepf, {\it Phys. Rev. D}, {\bf 92}, 071301 (2015).

\bibitem{Ritus}  V. I. Ritus, 
{\it Sov. Phys. JETP}, {\bf  30}, 1181 (1970).

\bibitem{Ritus1972} V. I. Ritus, {\it Annals of Physics}, {\bf 69}, 555 (1972).

\bibitem{Narozhnyi1969} N. B. Narozhnyi, {\it Sov. Phys. JETP}, {\bf 28}, 371 (1969).

\bibitem{Karb3} F. Karbstein, H. Gies, M. Reuter,  and M. Zepf, 
{\it Phys. Rev. D}, {\it  92}, 071301 (2015).

\bibitem{Adler1971}  S. L. Adler, 
{\it Ann. Phys. (N.Y.)}, {\bf 67}, 599 (1971).

 \bibitem{PRDS} S. V. Bulanov,  P. V. Sasorov, F. Pegoraro, H. Kadlecova, S. S. Bulanov,
T. Zh. Esirkepov, N. N. Rosanov, and  G. Korn, {\it  Phys. Rev. D},  {\bf 101}, 016016 (2020).

\bibitem{Tamb}  A. Di Piazza, M. Tamburini, S. Meuren, and C. H. Keitel, 
{\it Phys. Rev. A},  {\bf 98}, 012134 (2018).

\bibitem{DiPiazza2019} A. Di Piazza, M. Tamburini, S. Meuren, and C. H. Keitel
{\it Phys. Rev. A}, {\bf 99}, 022125 (2019).

\bibitem{Ilderton2019a} A. Ilderton, {\it  Phys. Rev. D}, {\bf 100}, 125018 (2019).

\bibitem{Ilderton2019b} A. Ilderton, B. King,  and D. Seipt, {\it Phys. Rev. A}, {\bf 99}, 042121 (2019).

\bibitem{SAU31} F. Sauter, 
{\it Zeit. f\"ur Phys.}, {\bf 69}, 742 (1931).

\bibitem{KdV1885} D. J. Korteweg and G. de Vries, {\it Philos. Mag.}, {\bf 39}, 422 (1885).

\bibitem{KP1970} B. B. Kadomstev and V. I. Petviashvili, {\it Sov. Phys. Dokl.}, {\bf 15}, 539 (1970).

\bibitem{KP2018} G. Biondini and D. E. Pelinovsky, 
{\it Scholarpedia}, {\bf 3}, 6539 (2018). 

\bibitem{Pod1} B. Podolsky, {\it Phys. Rev.}, {\bf 62},  68  (1942).

\bibitem{Pod2}  B. Podolsky and C. Kikuchi, {\it Phys. Rev.},  {\bf 65},  228 (1944).

\bibitem{Pod3}  B. Podolsky and P. Schwed, {\it Rev. Mod. Phys.},  {\bf 20}  40 (1948).

\bibitem{BM} A. Barut and G. H.  Mullen,
{\it Ann. Phys.}, {\bf  20}, 203  (1962).

\bibitem{LW} 
T. D. Lee and G. C.  Wick, {\it Phys. Rev. D}, {\bf  2}, 1033  (1970).

\bibitem{TN} 
R. Turcati and M. J. Neves, {\it Advances in High Energy Physics},
{\bf 2014}, 153953, (2014).

\bibitem{BBMB} 
L. H. C. Borges, F. A. Barone, C. A. M. de Melo, and F. E Barone
{\it Nuclear Physics B}, {\bf 944}, 114634 (2019).

\bibitem{MSMS} 
M. Soljac\v \i\'c and M. Segev
{\it Phys. Rev. A}, {\bf  62}, 043817 (2020).

\bibitem{Gusynin1} V. P. Gusynin and I. A. Shovkovy, {\it Can. J. Phys.}, {\bf  74}, 282
(1996).

\bibitem{Gusynin2} V. P. Gusynin and I. A. Shovkovy, {\it  J. Math. Phys.},  {\bf 40},
5406 (1999).

\bibitem{ostr}  
M. Ostrogradski,  {\it Mem. Ac. St. Petersburg}, {\bf  6}, 385 (1850).

\bibitem{t-jC} 
T-J. Chen, M. Fasiello, E. A. Lim,  and A. J. Tolley,
{\it JCAP}, {\bf 02}, 042 (2013).

 \bibitem{effective} 
C. Grosse-Knetter, {\it Phys. Rev. D}, {\bf  49}, 6709(1994).

\bibitem{HKAD2019} H. Kadlecova, G. Korn, and S. V. Bulanov, 
 {\it  Phys. Rev. D},  {\bf 99}, 036002 (2019).

\bibitem{Hodo} F. Pegoraro and S. V. Bulanov, 
{\it Phys. Rev. D}, {\bf  100}, 036004 (2019).

\bibitem{Hodo-bis} F. Pegorar and S. V. Bulanov, 
{\it  Phys. Lett. A} {\bf 384}, 126064 (2020).

\bibitem{Shukla} P. K. Shukla, M. Marklund, D. D. Tskhakaya, and B. Eliasson,
{\it Phys. Plasmas},  {\bf 11}, 3767 (2004).

 \bibitem{Mark} M.  Marklund and  J. Lundin, {\it Eur. Phys. J. D}, {\bf 55}, 319 (2009).

\bibitem{harm}  P. Sasorov, T. Zh. Esirkepov, F. Pegoraro, and S. V. Bulanov  (to be published).

\bibitem{WHIT1974} G. B. Whitham, {\it Linear and Nonlinear Waves} (Wiley, New York, 1974).

\bibitem{SPN1984} S. P. Novikov, S. V. Manakov, L. P. Pitaevskii, and V. E. Zakharov, {\it Theory of Solitons: The Inverse Scattering Method} (Springer-Verlag. 1984).

\bibitem{compact} P. Rosenau and J. M. Hyman, {\it Phys. Rev. Lett.}, {\bf 70},  564 (1993).



\end{thebibliography}
 \end{document}